\documentclass[aps,prd,twocolumn,superscriptaddress,nofootinbib]{revtex4-2}
\newcommand{\ibar}{{\declareslashed{}{\text{-}}{0.04}{0}{I}\slashed{I}}}
\newcommand{\jcap}{J. Cosmol. Astropart. Phys. }

\newcommand{\apjl}{Astrophys. J. Lett.}
\newcommand{\physrep}{Phys. Rep. }
\newcommand{\aap}{Astron. Astrophys. }

\newcommand{\mnras}{Mon. Not. R. Astron. Soc. }

\newcommand{\nphysa}{Nucl. Phys. A }
\usepackage{color}
\usepackage{amsmath}

\usepackage{graphicx}
\usepackage{slashed}
\newcommand{\eda}{\mathcal{E}}
\newcommand{\fla}{\mathcal{F}}

\begin{document}

\title{Multi-messenger signals of heavy axionlike particles in core-collapse supernovae: two-dimensional simulations}

\author{Kanji Mori}
\email[]{kanji.mori@nao.ac.jp}
\altaffiliation{Research Fellow of Japan Society for the Promotion of Science}
\affiliation{National Astronomical Observatory of Japan, 2-21-1 Osawa, Mitaka, Tokyo 181-8588, Japan}
\affiliation{Research Institute of Stellar Explosive Phenomena, Fukuoka University, 8-19-1 Nanakuma, Jonan-ku, Fukuoka-shi, Fukuoka 814-0180, Japan}
\author{Tomoya Takiwaki}
\affiliation{National Astronomical Observatory of Japan, 2-21-1 Osawa, Mitaka, Tokyo 181-8588, Japan}
\author{Kei Kotake}
\affiliation{Research Institute of Stellar Explosive Phenomena, Fukuoka University, 8-19-1 Nanakuma, Jonan-ku, Fukuoka-shi, Fukuoka 814-0180, Japan}
\affiliation{Department of Applied Physics, Faculty of Science, Fukuoka University, 8-19-1 Nanakuma, Jonan-ku, Fukuoka-shi, Fukuoka 814-0180, Japan}
\affiliation{Institute for Theoretical Physics, University of Wroc\l aw, 50-204 Wroc\l aw, Poland}
\author{Shunsaku Horiuchi}
\affiliation{Center for Neutrino Physics, Department of Physics, Virginia Tech, Blacksburg, VA 24061, USA}
\affiliation{Kavli IPMU (WPI), UTIAS, The University of Tokyo, Kashiwa, Chiba 277-8583, Japan}

\date{\today}

\begin{abstract}
Core-collapse supernovae are a useful laboratory to probe the nature of exotic particles. If axionlike particles (ALPs) are produced in supernovae, they can affect the transfer of energy and leave traces in observational signatures. In this work, we present results from two-dimensional supernova models including the effects of the production and the absorption of ALPs that couple with photons. It is found that the additional heating induced by ALPs can enhance the diagnostic energy of explosion, $E_\mathrm{diag}$. For example, for moderate ALP-photon coupling, we find explosion energies $\sim0.6\times10^{51}$\,erg compared to our reference model without ALPs of $\sim0.4\times10^{51}$\,erg in the first $\sim0.5$\,s postbounce explored in this work. Our findings  indicate that when the coupling constant is sufficiently high, the neutrino luminosities and mean energies are decreased because of the additional cooling of the proto-neutron star via ALPs. The gravitational wave amplitude is also reduced because the mass accretion on the proto-neutron star is suppressed. Although the ALP-photon coupling can foster explodability, including enhancing the explosion energy closer to recent observations, more long-term simulations in spatially three-dimension are needed to draw robust conclusions. 

\end{abstract}

\maketitle

\section{Introduction}

Core-collapse supernovae are a major target of multi-messenger astronomy. In 1987, electron anti-neutrinos from SN 1987A, which appeared in the Large Magellanic  Cloud, were detected by the IMB \cite{1987PhRvL..58.1494B}, Kamiokande \cite{1987PhRvL..58.1490H}, and Baksan \cite{1987JETPL..45..589A} experiments. As a result, the prolific emission of neutrinos from stellar core collapse was experimentally verified. Also, several gravitational wave (GW) detectors, namely LIGO, VIRGO, and KAGRA, have started their operation and are waiting for signals from a nearby supernova \cite[e.g.,][]{2021PhRvD.104l2004A}. Neutrinos and GWs interact with matter so feebly that they can provide information on the core of collapsing stars \cite{2013CRPhy..14..318K,2016NCimR..39....1M,2018JPhG...45d3002H}. In order to extract such information from multi-messenger signals, it is necessary to develop realistic supernova models.

In the core of collapsing stars,  temperatures of $\sim10$\,MeV and densities of $\sim10^{14}$\,g\,cm$^{-3}$ are reached. Such extreme environments in supernovae are useful to probe exotic physics beyond the Standard Model \cite[e.g.,][]{1996slfp.book.....R,2018JPhG...45d3002H,2022JCAP...12..024B}. For example, the effect of the production of axionlike particles (ALPs;  \cite{1978PhRvL..40..223W,1978PhRvL..40..279W,1978JETPL..27..502V,doi:10.1146/annurev-nucl-120720-031147}) that interact with photons has  gained a lot of attention recently and been investigated theoretically. If such new particles are produced in a proto-neutron star (PNS), they can induce an additional energy loss. As a result, the duration of neutrino signals from supernova events can be shorter than the observed duration for SN 1987A. This argument has been adopted to obtain constraints on the ALP mass and the coupling constant \cite{1988PhRvL..60.1797T,1988PhRvL..60.1793R,1995PhRvD..52.1755M,2018arXiv180810136L,2020JCAP...12..008L,2022arXiv220914318F}. Once produced, ALPs can also decay during their propagation through the stellar envelope. In this case, ALPs work as an additional heating source and can lead to more energetic explosions than standard supernova models \cite{1982ApJ...260..868S,2019PhRvD..99l1305S,2022PhRvD.105f3009M,2022PhRvL.128v1103C,2022PhRvD.105c5022C}. If the mean free path of ALPs is longer than the stellar radius, they decay outside the envelope and produce $\gamma$-rays which may be observed by space telescopes. Non-detection of $\gamma$-rays from SN 1987A have provided constraints on ALPs \cite{2011JCAP...01..015G,2015JCAP...02..006P,2018PhRvD..98e5032J,2022arXiv220513549B,2023arXiv230311395D,2023arXiv230510327D}, and future observations of a nearby supernova will provide more stringent constraints \cite{2017PhRvL.118a1103M,2021PhRvL.127r1102C,2022PhRvD.105b3020M}. 

Most previous studies of ALPs in supernovae adopted a post-processing technique which decouples the ALP production from the hydrodynamics. However, stellar core-collapse models with coupled ALPs have recently been developed by several authors \cite{1982ApJ...260..868S,2016PhRvD..94h5012F,2021PhRvD.104j3012F,2022PhRvD.105f3009M,PhysRevD.106.063019}. Most of these coupled simulations have assumed spherical symmetry. The exception is the recent pioneering work \cite{PhysRevD.106.063019} which performed two-dimensional simulations with ALPs that couple with nucleons. However, ALPs can couple with other particles such as photons and electrons as well, and multi-dimensional models with these types of interactions have not yet been developed. In particular, the ALP-photon interaction is interesting because it induces not only new cooling processes but also additional heating through ALP radiative decay.

One-dimensional models are computationally inexpensive, making them useful to investigate the dependence of supernova dynamics on the ALP parameters such as mass and coupling constant. However, in general, the accretion onto the PNS deviates from spherically symmetric flows. Hydrodynamical instabilities such as convection and the standing accretion shock instability (SASI; \cite{2002A&A...392..353F,2003ApJ...584..971B}) are thought to play vital roles in reenergizing the stalled shock
into expansion. To track these effects, it is necessary to investigate the effect of ALPs using multi-dimensional models. Also, spherically-symmetric models cannot predict GWs, which can provide additional multi-messenger observational probes of the supernova core. In this study, we develop such two-dimensional axisymmetric supernova models with ALPs that couple with photons. 

This paper is organized as follows. In Section II, the computational setup is described. In Section III, we show the result of our simulations, including explosion properties and neutrino and GW signals. In Section IV, we discuss implications of our results. 

 \section{Method}
 \label{sec:method}

\begin{table*}[]
\begin{tabular}{ccc|ccc}
Model&$m_a$ [MeV] & $g_{10}$ & $t_\mathrm{pb,\;2000}$ [ms]& $E_\mathrm{diag}$ {[}$10^{51}$ erg{]} &$M_\mathrm{PNS}/M_\odot$\\\hline\hline
   Standard&$-$   & 0  &   390   & 0.40 &    1.78                                \\\hline

$(100,\;2)$&100   & 2        &   385         &   0.37                                   & 1.77                \\
$(100,\;4)$&100   & 4        &    362                                 &  0.34&  1.76                   \\
$(100,\;10)$&100   & 10           &395                                       &0.36&    1.77                  \\
$(100,\;12)$&100   & 12           &357      & 0.43 &   1.77                     \\
$(100,\;14)$&100   & 14        &   360         &   0.44                                  &    1.77               \\
$(100,\;16)$&100   & 16        &    367                                 &  0.51    &1.77               \\
$(100,\;20)$&100   & 20           &330                                       &1.10&  1.74                                       \\\hline
$(200,\;2)$&200   & 2        &   374         &   0.45                                  &    1.77               \\
$(200,\;4)$&200   & 4        &    376                                 &  0.45&    1.76            \\
$(200,\;6)$&200   & 6           &333                                       &0.54&           1.75           \\
$(200,\;8)$&200   & 8           &323      & 0.94&  1.74                     \\
$(200,\;10)$&200   & 10        &   319         &   1.61                                  &   1.73                \\
$(200,\;20)$&200   & 20        &    248                                 &  3.87&     1.62               \\

\end{tabular}
\caption{The supernova models developed in this work. The row with $g_{10}=g_{a\gamma}/10^{-10}\,\mathrm{GeV}^{-1}=0$ represents the 
model without ALPs. Each model with ALPs is designated by a pair of two numbers which represents  $(m_a/1\,\mathrm{MeV},\;g_{10})$, where $m_a$ is the ALP mass and $g_{10}$ is the ALP-photon coupling constant. The fourth column shows $t_\mathrm{pb,\;2000}$, which is the post-bounce time at the moment when the shock  wave reaches $r=2000$ km. The fifth column shows the diagnostic explosion energy $E_\mathrm{diag}$ at $t_\mathrm{pb}=t_{\mathrm{pb},\;2000}$. The last column shows the PNS mass $M_\mathrm{PNS}$ evaluated at  $t_\mathrm{pb}=t_\mathrm{pb,\;2000}$.}
\end{table*}

In this work, we closely follow the method adopted in the one-dimensional spherical simulations in Ref.~\cite{2022PhRvD.105f3009M} except for the spatial dimensionality. In this section, we briefly describe the ALP models and the simulation setup.
 
The ALP-photon interaction is described by the Lagrangian \citep{1988PhRvD..37.1237R}
 \begin{equation} 
    \mathcal{L}=-\frac{1}{4}g_{a\gamma}F_{\mu\nu}\tilde{F}^{\mu\nu}a,\label{lag}
\end{equation}
where $g_{a\gamma}$ is the coupling constant, $F_{\mu\nu}$ is the electromagnetic tensor, and $a$ is the ALP field. This interaction induces the Primakoff process $(\gamma+p\rightarrow a+p)$ and photon coalescence $(\gamma+\gamma\rightarrow a)$ which produce ALPs from photons in the plasma. We implement these two processes as ALP production processes. Our prescription for the ALP production rate, $Q_\mathrm{cool}$, is given in Ref.~\citep{2022PhRvD.105f3009M}. Also, the inverse Primakoff process $(a+p\rightarrow \gamma +p)$ and radiative decay $(a\rightarrow \gamma+\gamma)$ are implemented to calculate ALP heating. Our prescription for the ALP absorption rate, $Q_\mathrm{heat}$, is also given in Ref.~\citep{2022PhRvD.105f3009M}. We adopt ALP masses of $m_a=100$ and 200\,MeV and ALP-photon coupling constants of $g_{10}=g_{a\gamma}/10^{-10}\,\mathrm{GeV}^{-1}=2$--$20$. We focus on this parameter range because one-dimensional core-collapse simulations performed in Ref.~\citep{2022PhRvD.105f3009M} indicate that the revival of the stalled shock can be assisted by the additional ALP heating in this parameter space. Although comparisons between supernova models and observed low-energy supernovae exclude most of this ALP parameter range \cite{2022PhRvL.128v1103C}, we adopt these parameters to demonstrate that stellar core-collapse simulations predict signatures of exotic particles in observable multi-messenger signals and furthermore our findings motivate further studies focusing on other exotic physics. We also develop and compare with a model without ALPs. 

We implement the ALP processes above in the supernova simulation code \texttt{3DnSNe} \cite{2016MNRAS.461L.112T}. We perform two-dimensional core-collapse simulations with spatial resolution $n_r\times n_\theta=512\times 128$ and simulate out to radius 5000\,km. The nuclear equation of state is from Ref.~\cite{1991NuPhA.535..331L} with $K=220$\,MeV. We adopt the three-flavor isotropic diffusion source approximation for neutrino transport \cite{2009ApJ...698.1174L,2014ApJ...786...83T,2018ApJ...853..170K}. We use the non-rotating $20M_\odot$ solar metallicity progenitor model from Ref.~\cite{2007PhR...442..269W}.

In order to treat the ALP transport, we start from the zeroth angular moment of the Boltzmann equation
\begin{eqnarray}
\frac{\partial \eda }{\partial t}
+\nabla\cdot \mathbf{\fla}
=Q_\mathrm{cool}-Q_\mathrm{heat},
\end{eqnarray}
where $\eda$ is the ALP energy per unit volume and $\mathbf{\fla}$ the ALP energy flux. We drop the term $\partial \eda/\partial t$, assuming the stationarity of the ALP flux. We then adopt the ray-by-ray approximation, in which ALPs are assumed to propagate only in the radial direction. The ALP luminosity, $L_{\rm ALP}$, that is defined at the edges of $i$-th radial cell, follows from the relation
 \begin{eqnarray}
 L_{{\rm ALP},i+\frac{1}{2}}=L_{{\rm ALP},i-\frac{1}{2}}+(Q_{\mathrm{cool},\;i}-Q_{\mathrm{heat},\;i})\Delta V_i,\label{rec}
 \end{eqnarray}
where $Q_{\mathrm{cool},\;i}$ and $Q_{\mathrm{heat},\;i}$ are the ALP cooling and heating rates and $\Delta V_i$ is the volume of the $i$-th cell. This relation is coupled with
 \begin{eqnarray}
 Q_{\mathrm{heat},\;i}\Delta V_i=L_{{\rm ALP},i-\frac{1}{2}}\left(1-\exp\left(-\frac{r_{i+1}-r_{i}}{\lambda_{a,\;i}}\right)\right),\label{qheat}
 \end{eqnarray}
 which determines $Q_{\mathrm{heat},\;i}$. Solving Eqs.~(\ref{rec}) and (\ref{qheat}), we can obtain values of $Q_{\mathrm{heat},\;i}$ and  $Q_{\mathrm{heat},\;i}$ for every $i$. At the $n$-th time step, ALPs are  coupled with hydrodynamics as 
 \begin{eqnarray}
e_{\mathrm{int},\;i}^{n+1} = e_{\mathrm{int},\;i}^n +(Q_{\mathrm{heat},\;i}^n -Q_{\mathrm{cool},\;i}^n)\Delta t,
\end{eqnarray}
where $e_{\mathrm{int},\;i}$ is the internal energy and $\Delta t$ is the time step size.

 \begin{figure}
  \centering
  \includegraphics[width=8cm]{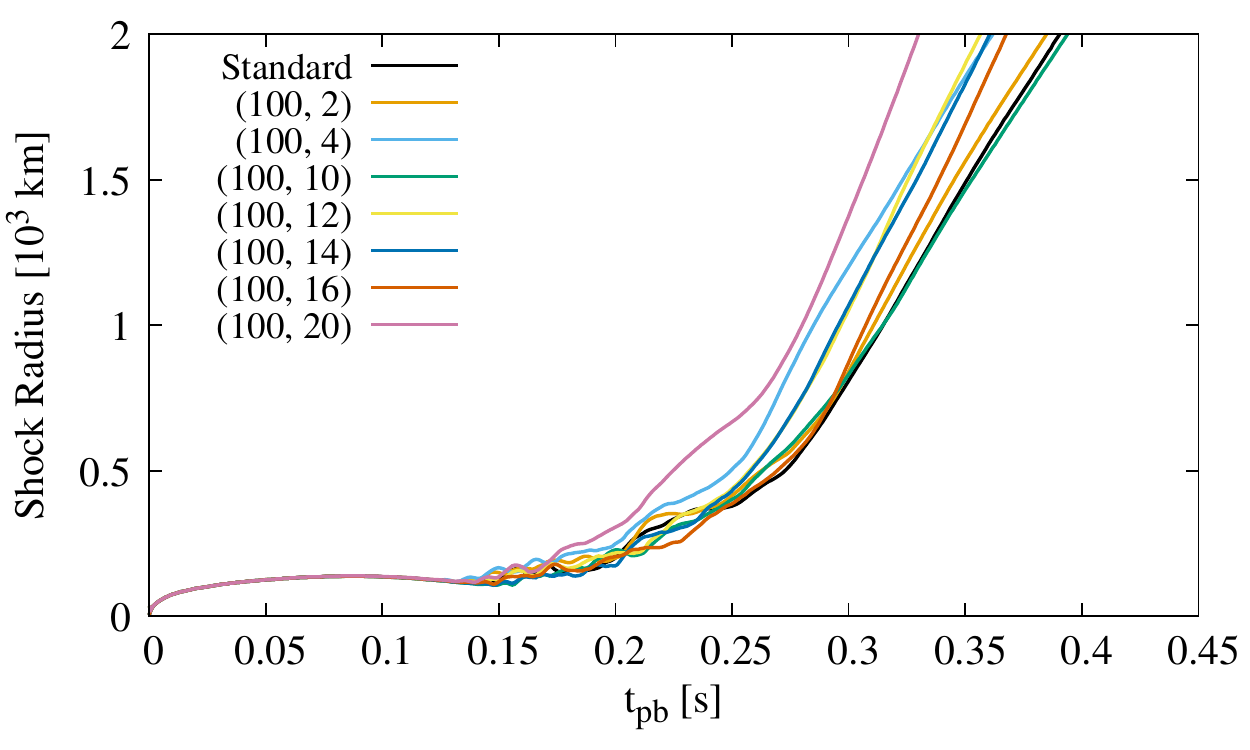}
   \includegraphics[width=8cm]{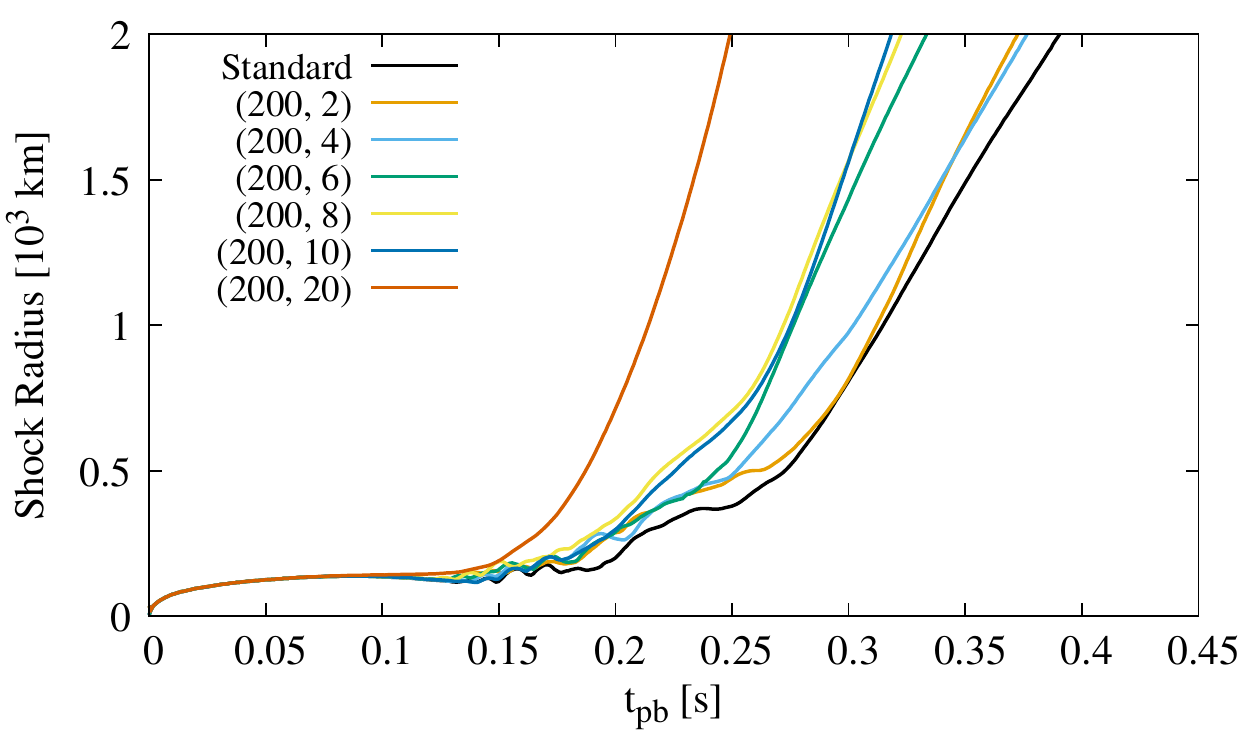}
  \caption{The average radius of the bounce shock as a function of the time $t_\mathrm{pb}$ after the core bounce. The upper panel shows the results for the models with $m_a=100$\,MeV and the lower panel is for the models with $m_a=200$\,MeV. The pairs of integers in the legend indicate $(m_a/1\,\mathrm{MeV},\;g_{10})$, and this notation is used throughout this paper.}
  \label{rsh}
 \end{figure}

\section{Results}
In this work, we develop 13 models with ALPs and one model without ALPs, as tabulated in Table 1. In this section, we describe the properties of these models.

\subsection{Explosion Properties}
\begin{figure}
  \centering
  \includegraphics[width=8cm]{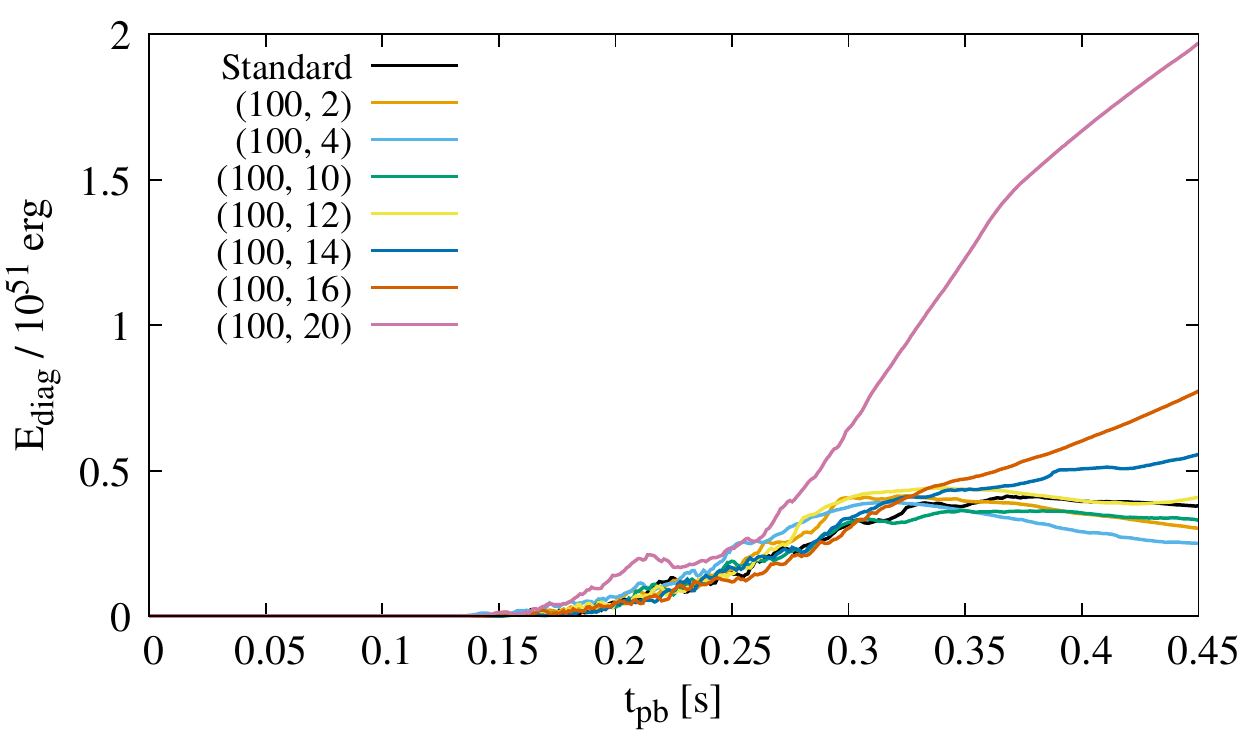}
   \includegraphics[width=8cm]{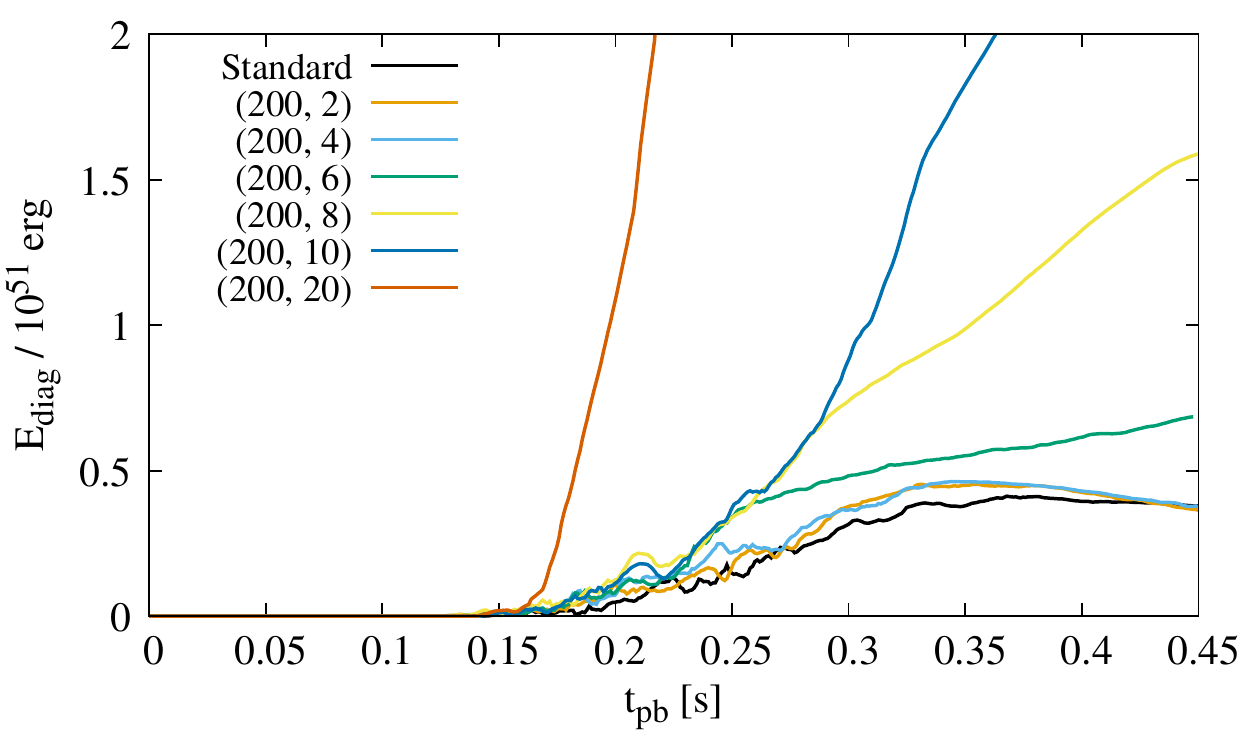}
  \caption{The diagnostic explosion energy $E_\mathrm{diag}$ as a function of the time $t_\mathrm{pb}$ after the core bounce. The upper panel shows the results for the models with $m_a=100$\,MeV and the lower panel is for the models with $m_a=200$\,MeV. }
  \label{Eexp}
\end{figure}

  \begin{figure*}
  \centering
  \includegraphics[width=5.5cm]{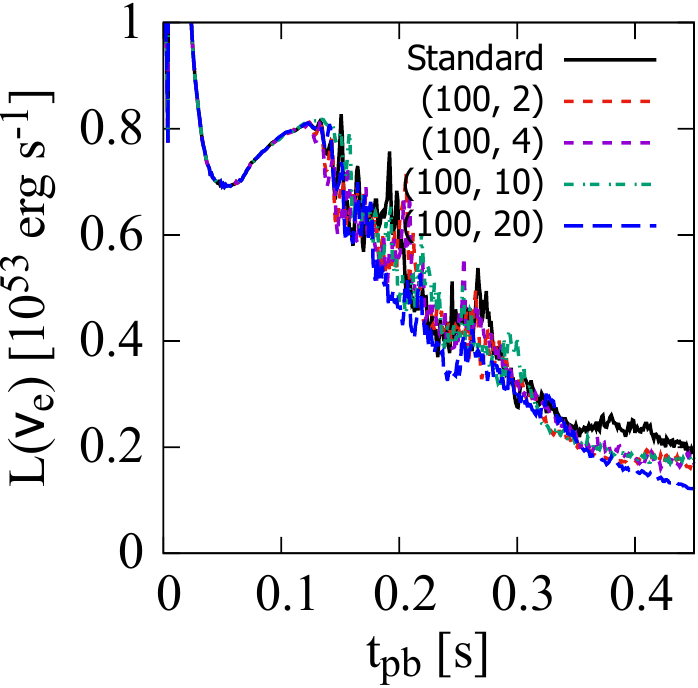}
   \includegraphics[width=5.5cm]{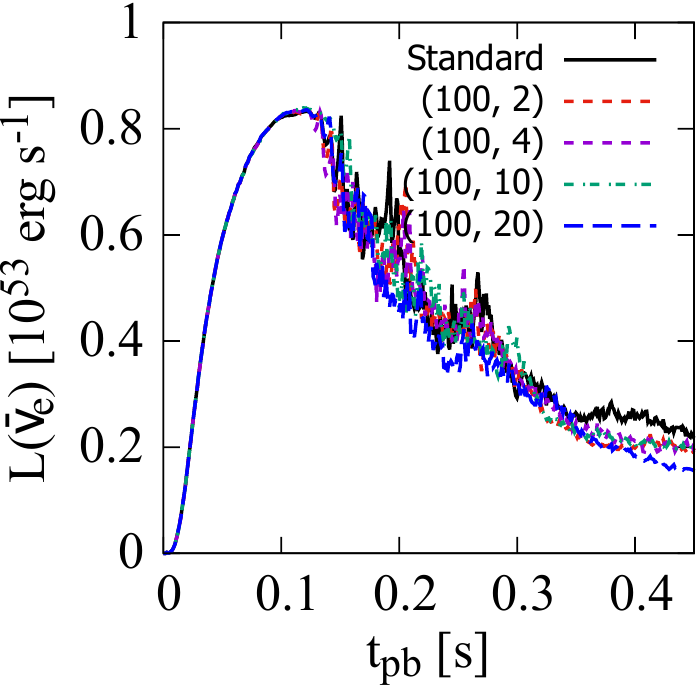}
   \includegraphics[width=5.5cm]{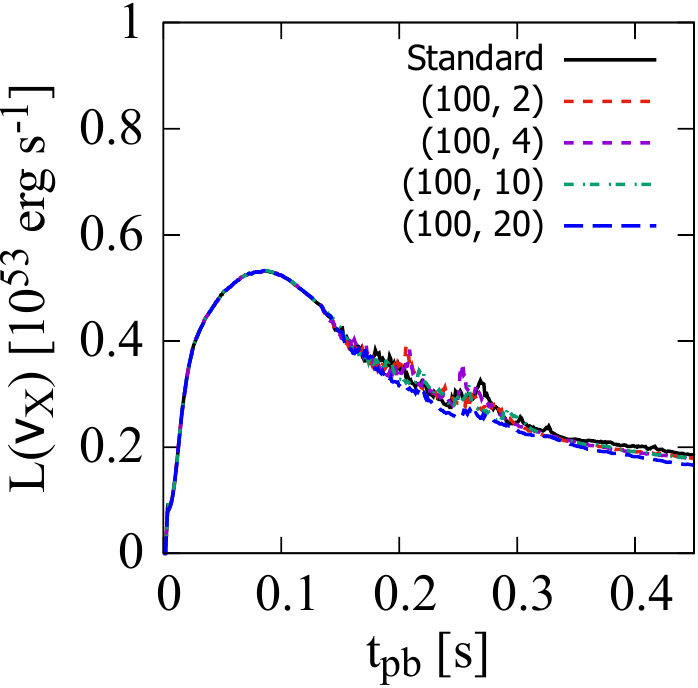}
   \includegraphics[width=5.5cm]{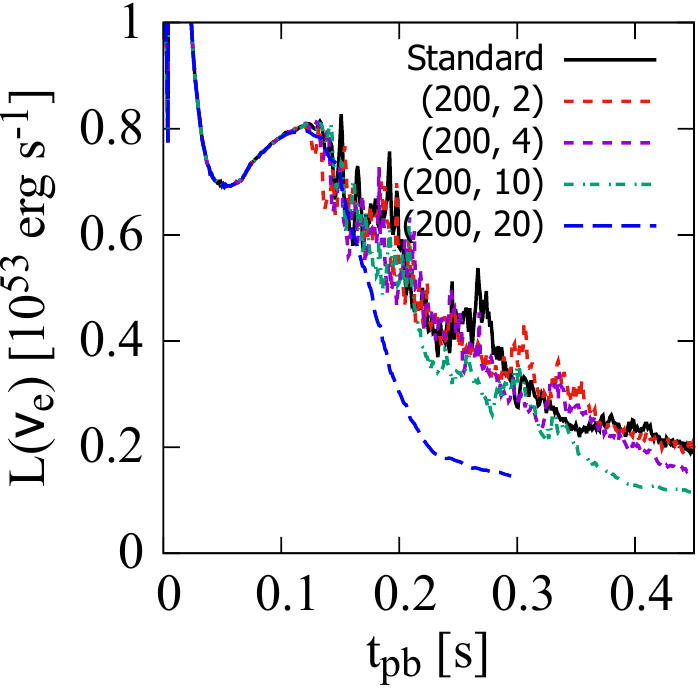}
   \includegraphics[width=5.5cm]{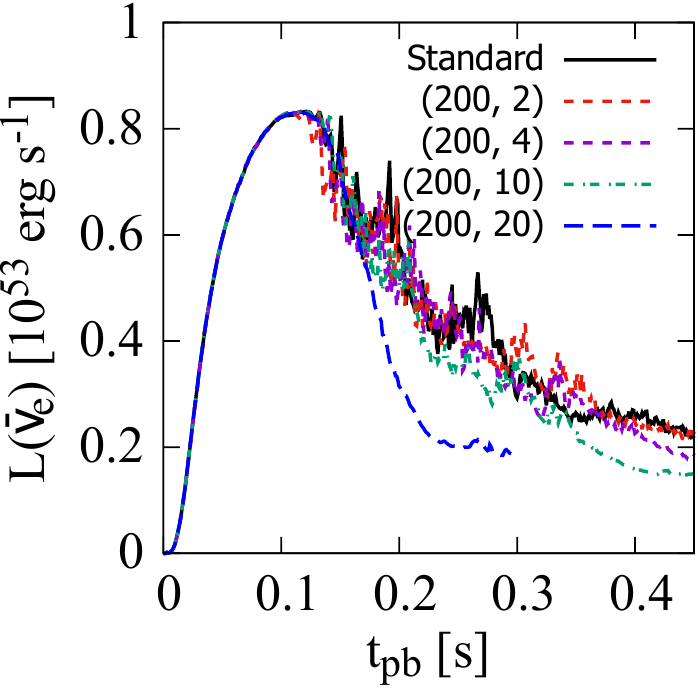}
   \includegraphics[width=5.5cm]{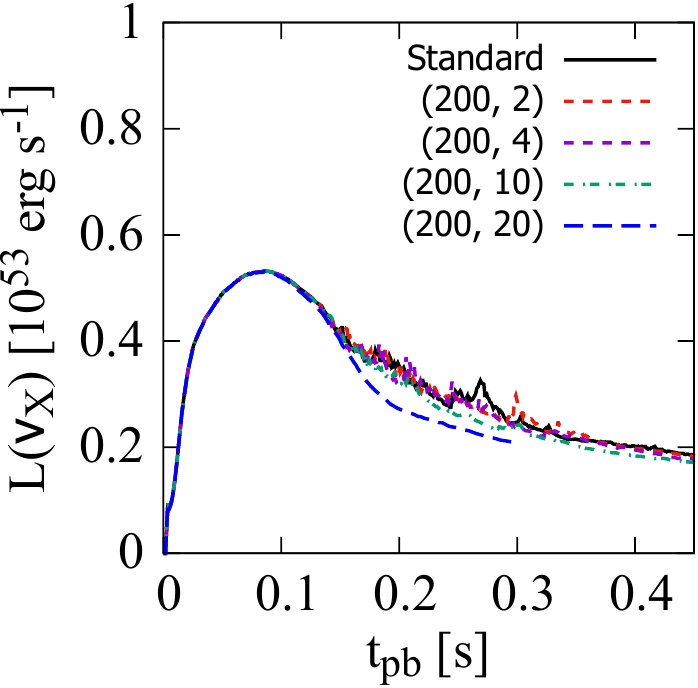}
  \caption{The luminosities of $\nu_e$, $\bar{\nu}_e$, and $\nu_X$ in the selected models as a function of the time $t_\mathrm{pb}$ after the core bounce. The solid line corresponds to the model without ALPs, and the other lines correspond to the models with ALPs.}
  \label{Ln}
 \end{figure*}
   \begin{figure*}
  \centering
  \includegraphics[width=5.5cm]{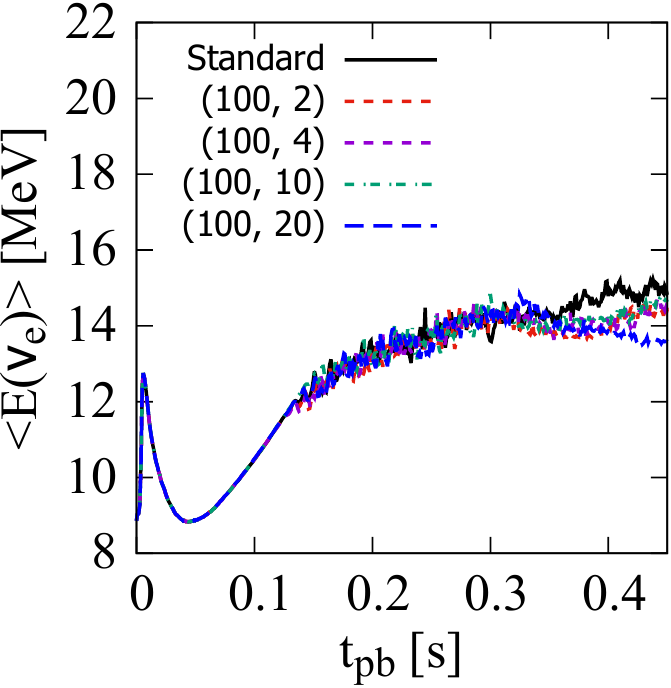}
   \includegraphics[width=5.5cm]{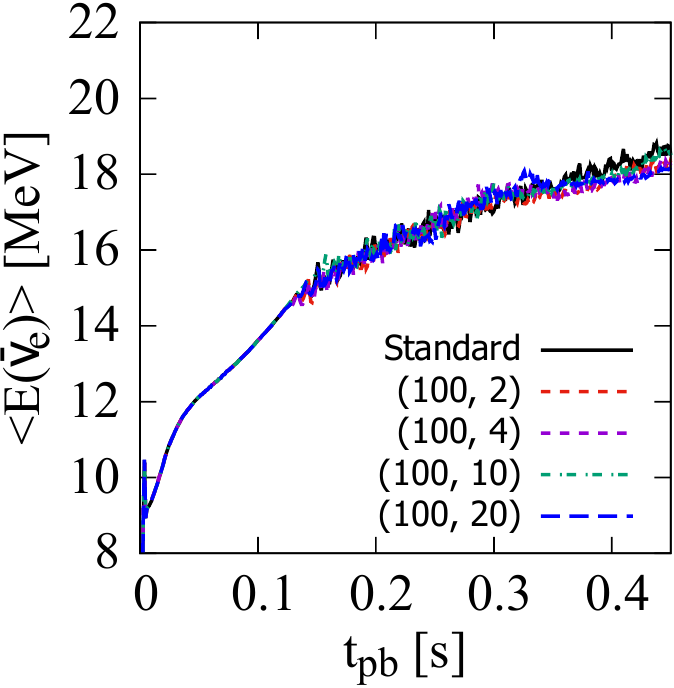}
   \includegraphics[width=5.5cm]{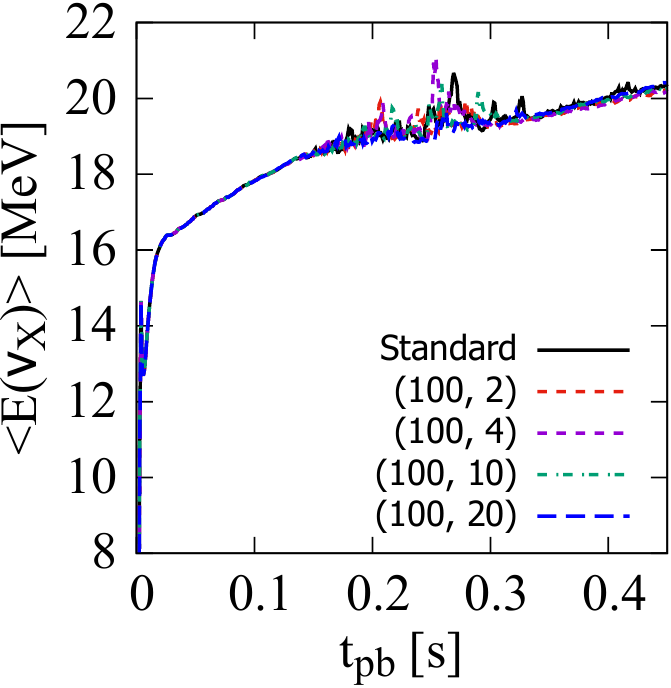}
   \includegraphics[width=5.5cm]{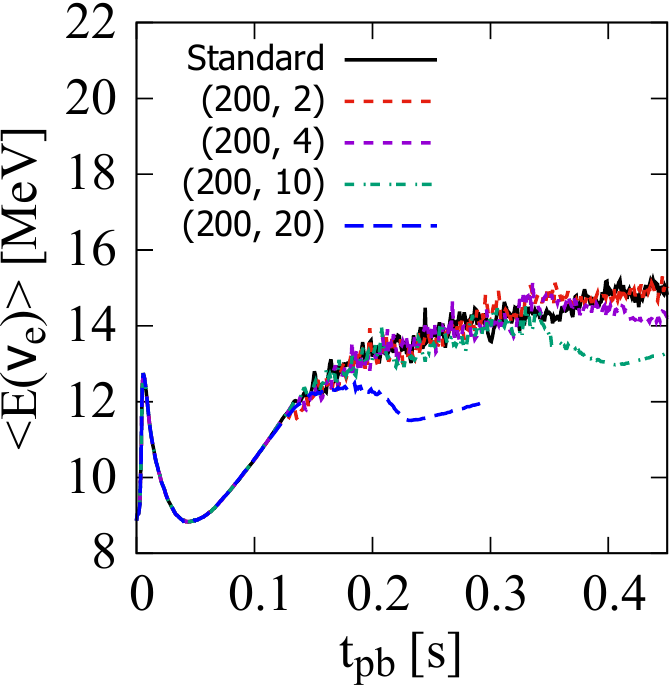}
   \includegraphics[width=5.5cm]{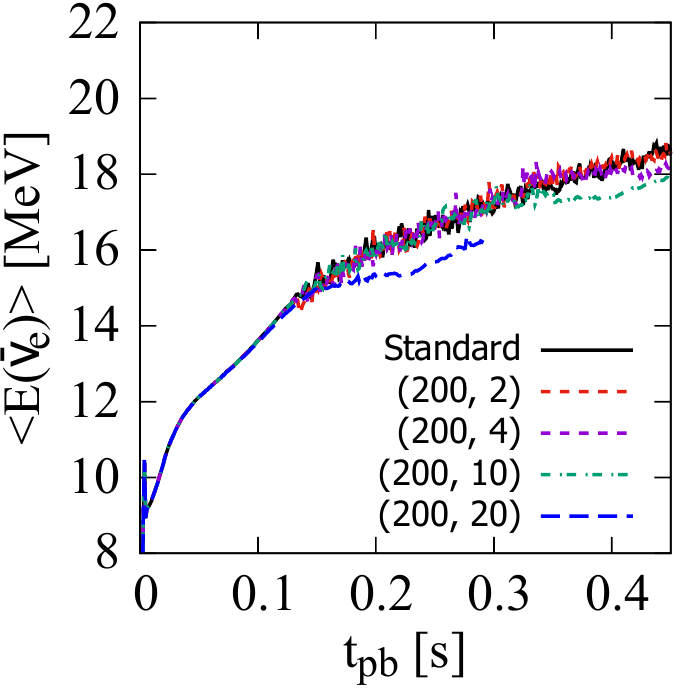}
   \includegraphics[width=5.5cm]{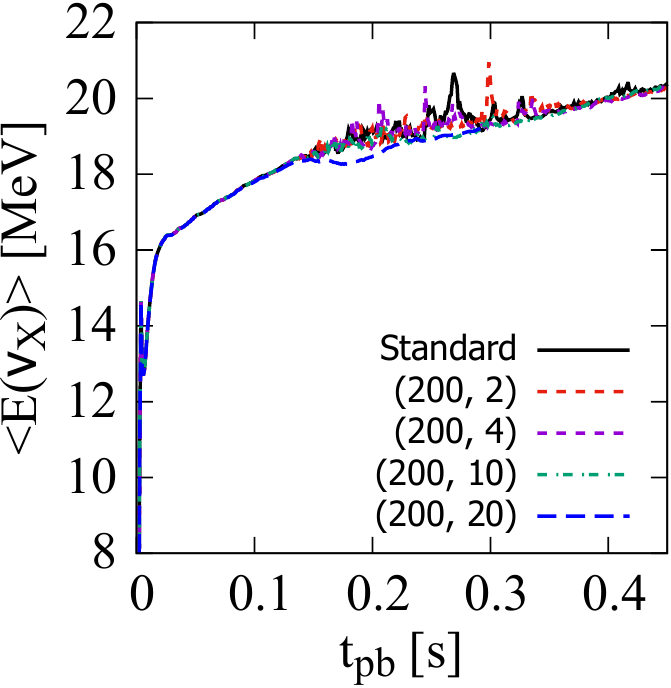}
  \caption{The mean energy of $\nu_e$, $\bar{\nu}_e$, and $\nu_X$ in the selected models as a function of the time $t_\mathrm{pb}$ after the core bounce. The solid line corresponds to the model without ALPs, and the other lines correspond to the models with ALPs.}
  \label{En}
 \end{figure*}
 
When a massive star reaches the end of its life, its iron core starts collapsing and the central density increases. The density becomes reaches the nuclear saturation density, and the equation of state stiffens. This causes core bounce, which leads to the formation of the bounce shock. Although the shock initially stalls, it can be pushed outward (revived) because of neutrino and ALP heating, and the supernova becomes optically luminous when the shock wave passes the stellar photosphere.

The success of a supernova explosion depends on whether the shock wave is revived or not. In one-dimensional models, the shock wave is typically not revived and the explosion  fails \cite[e.g.,][]{2018JPhG...45j4001O}; the exception is the lightest stars \cite{2006A&A...450..345K,2008A&A...485..199J}. Multi-dimensional models can typically successfully explode, although explodability is not a simple function of the stellar mass \cite[e.g.,][]{2014ApJ...783...10S}.

Figure~\ref{rsh} shows the averaged radius of the bounce shock. The solid curve shows the standard model without ALPs and the other curves show the models with ALPs. Because they are two-dimensional models, the shock wave is revived even if only neutrino heating operates, i.e., even without ALPs. However, the figure shows that ALP heating makes the shock radius grow larger faster. In the one-dimensional models of Ref.~\citep{2022PhRvD.105f3009M}, the shock wave was revived only if $g_\mathrm{a\gamma}$ was larger than some critical value with shock revival occurring earlier when $g_{a\gamma}$ was larger. The result for our two-dimensional models is  similar to the one-dimensional models, except that the additional heating is not necessary for shock revival.

Figure \ref{Eexp} shows the diagnostic explosion energy defined as
   \begin{eqnarray}
  E_\mathrm{diag}=\int_D dV\left(\frac{1}{2}\rho v^2+e-\rho\Phi\right),
  \end{eqnarray}
where $\rho$ is the density, $v$ is the fluid velocity, $e$ is the internal energy, $\Phi$ is the gravitational potential, and $D$ is the region where the total energy is positive and the radial velocity is outward\footnote{The condition $v_r>0$, where $v_r$ is the radial velocity, is sometimes omitted from the definition of the diagnostic energy of explosion \cite[e.g.][]{2015ApJ...808L..42M,2019MNRAS.485.3153B}. The value of $E_\mathrm{diag}$  with our definition is smaller than that with the other definition. The quantitative and systematic comparison between these definitions is out of the scope of this study, but it is desirable to perform such a study even in the standard framework without ALPs.}. It is seen that the explosion energy for the standard model saturates at $E_\mathrm{diag}\approx0.4\times10^{51}$\,erg. However, when ALPs are included, $E_\mathrm{diag}$ tends to increase with $g_{a\gamma}$ because of higher heating rates. In the models with $m_a=100$\,MeV and $g_{10}\leq 10$, however, $E_\mathrm{diag}$ is smaller than the standard model at $t_\mathrm{pb}\sim0.45$\,s. This could be attributed to stochasticity of the turbulent motion. It is notable that two-dimensional supernova models with axion-nucleon coupling show similar non-monotonic behavior when the coupling constant is relatively small \cite{PhysRevD.106.063019}. Also, it can be seen that heavier ALPs lead to more energetic explosion with a fixed $g_\mathrm{a\gamma}$. This is because the mean free path of heavier ALPs is shorter and thus heat the gain region behind the shock more efficiently.

In Fig.~\ref{Eexp}, we can see that $E_\mathrm{diag}$ becomes higher when ALP heating is considered. In the models with $g_{10}\geq16$ for $m_a=100$\,MeV and $g_{10}\geq6$ for $m_a=200$\,MeV,  $E_\mathrm{diag}$ reaches $0.6\times10^{51}$\,erg and is still growing at the end of simulation. In particular, the $(200,\;6)$ model is an interesting case. In this model, the explosion energy is higher than that in the standard model without ALPs by $\sim0.2\times10^{51}$\,erg at the end of simulation. The resultant energy is closer to the observed values \cite{2022A&A...660A..41M}. Although it is excluded using low-energy supernovae \cite{2022PhRvL.128v1103C}, the ALP parameter is close to the edge of the excluded parameter region. In Ref.~\cite{2022PhRvL.128v1103C}, they adopted a supernova model with the progenitor mass $18.8M_\odot$ to obtain their constraint. However, such low-energy supernovae would be originated from lighter progenitors including low-mass iron core stars and super-asymptotic giant branch stars \cite[e.g.][]{2020MNRAS.496.2039S}. Since the cooler cores formed in the lighter stars would lead to lower ALP luminosities, the constraint on ALPs may be relaxed. It is hence desirable to perform detailed studies on the progenitor dependence. Observationally, light curve modeling of supernova events implies that the most frequent value of the type II supernova explosion energy is $E_\mathrm{exp}\sim0.6\times10^{51}$\,erg \cite{2022A&A...660A..41M}. Also, a detailed analysis with three-dimensional supernova models estimates the explosion energy of SN 1987A as $E_\mathrm{exp}\approx1.5\times10^{51}$\,erg \cite{2020MNRAS.494.2471J}. We cannot directly compare these observational values with the models because the diagnostic energy likely increases even after the simulated time range. Nevertheless, we can see that the ALP heating can render the explosion energy higher. 

We also estimate the explosion energy that consider the overburden of the unshocked region \cite{2013ApJ...767L...6B,2021ApJ...915...28B}
\begin{eqnarray}
E_\mathrm{OB}=E_\mathrm{diag}+\int_{M_r(r_\mathrm{sh})}^{M_r(r_\mathrm{surf})}\left(e-\frac{GM_r}{r}\right)dM_r.
\end{eqnarray}
Here, $M_r$ is the mass coordinate, $r_\mathrm{sh}$ is the shock radius, $r_\mathrm{surf}$ is the radius of the progenitor, $e$ is the internal energy density, $G$ is the gravitational constant. We find that $E_\mathrm{OB}$ is smaller than $E_\mathrm{diag}$ by $\sim(0.2$--$0.3)\times10^{51}$\,erg when the shock radius is $\sim2000$--5000\,km because of the binding energy in the unshocked region. However, it has been pointed out that the effect of the overburden would be compensated by the energy release of nuclear processes in the accreted material \cite{2009ApJ...694..664M}. Also, we do not compare the absolute value of the explosion energy in each model with observations but focus on the differences between the models. We therefore use $E_\mathrm{diag}$ instead of $E_\mathrm{OB}$ to evaluate the explosion energy.

In our simulations, the region outside $r=5000$\,km is not included. However, ALPs can decay and deposit energy outside the simulated region when their mean free path is long enough. This additional heating can contribute to the asymptotic kinetic energy of the ejecta. We can estimate the deposited energy as
\begin{eqnarray}
    E_a=\int^t_0dt_\mathrm{pb}\left.L_{\rm ALP}\right|_{r=5000\,{\rm km}}.
\end{eqnarray}
Here $\left.L_{\rm ALP}\right|_{r=5000\,{\rm km}}$ is the ALP luminosity at $r=5000\,{\rm km}$.
We found $E_a\sim0.1\times10^{51}$\,erg at the end of the simulations for the (100,\;2) and (200,\;2) models, which adopt coupling constants on the edge of the upper limit based on low-energy supernovae \cite{2022PhRvL.128v1103C}. This value of $E_a$ is consistent with the result reported in Ref.~\cite{2022PhRvL.128v1103C}. The higher $g_{a\gamma}$ is, the larger $E_a$ becomes. In particular, for the models with $g_{10}\geq 6$, $E_a$ exceeds $1\times10^{51}$\,erg at the end of the simulations. This implies that the explosion energy could exceed typical values for observed supernova events after the simulated time.
\subsection{Neutrinos}

    \begin{figure}
  \centering
  \includegraphics[width=8cm]{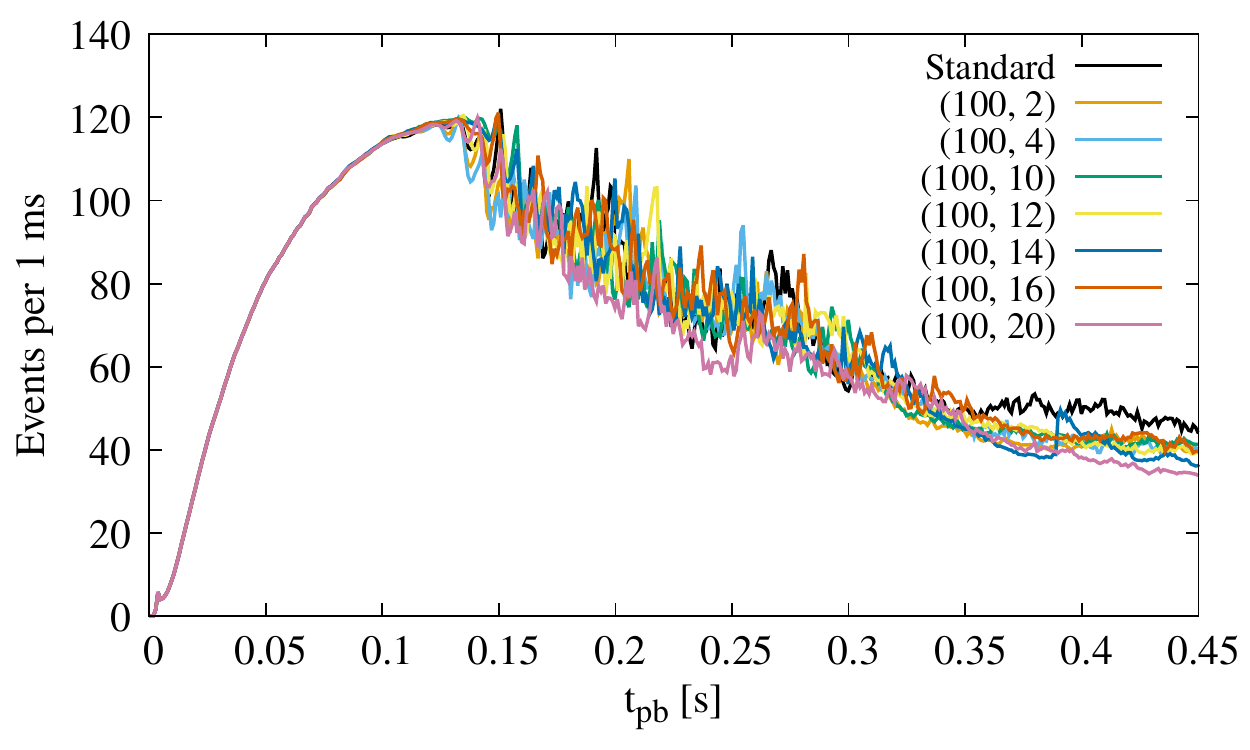}
   \includegraphics[width=8cm]{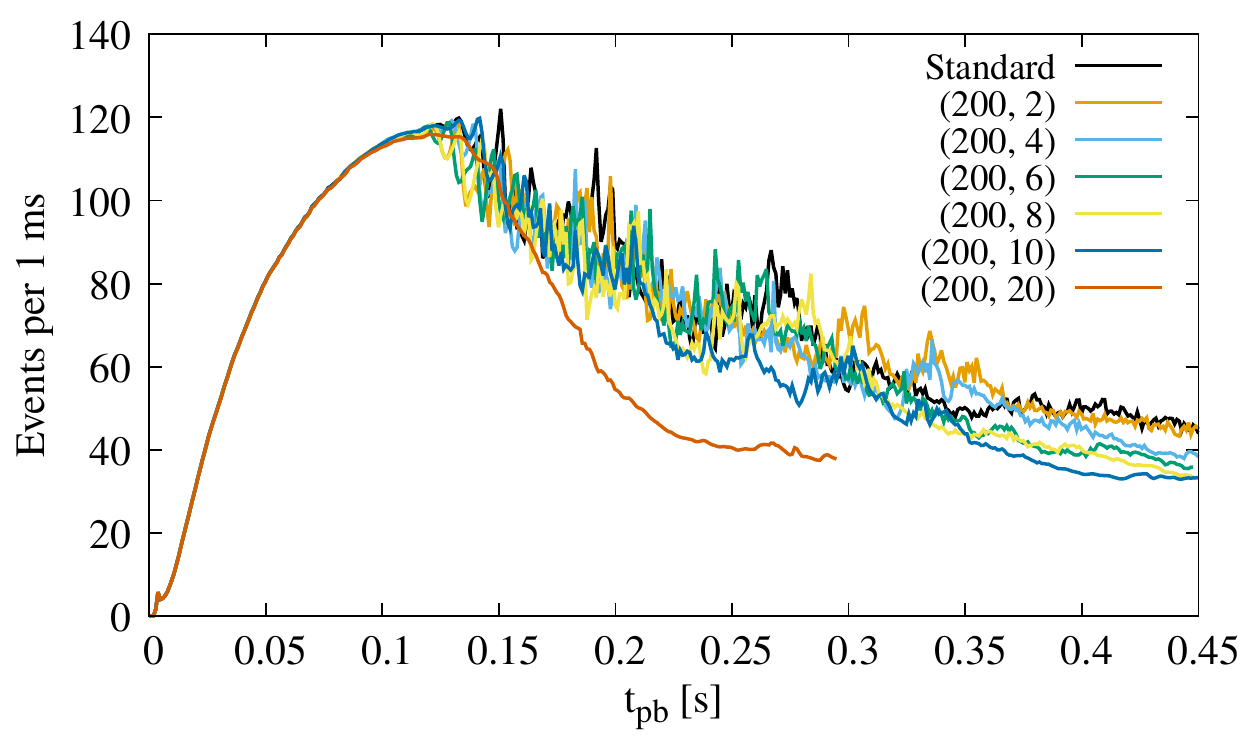}
  \caption{The number of neutrino events per 1\,ms from a supernova at the Galactic center detected by HK. The inverse $\beta$ decay is used to detect $\bar{\nu}_e$. The neutrino mass hierarchy is assumed to be normal and the distance to the supernova event is $D=8.5$\,kpc. The upper panel shows the results for the models with $m_a=100$\,MeV and the lower panel is for the models with $m_a=200$\,MeV.}
  \label{NH}
 \end{figure}

    \begin{figure}
  \centering
  \includegraphics[width=8cm]{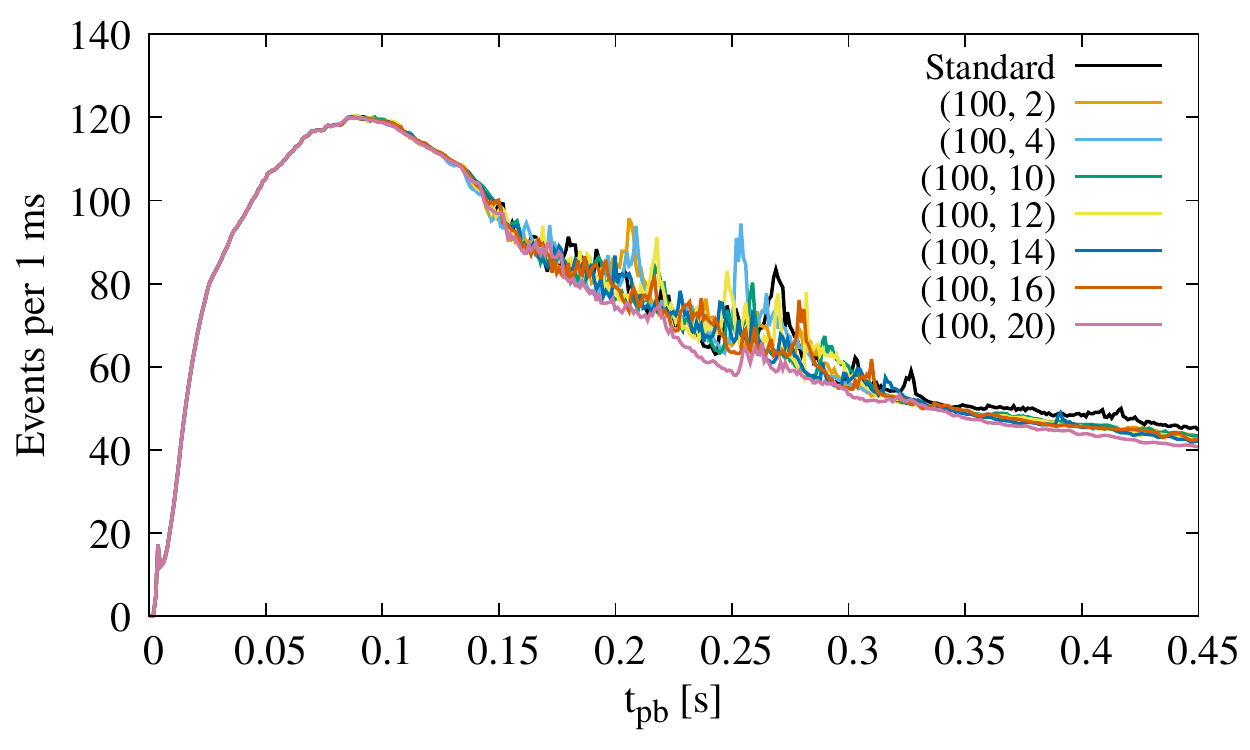}
   \includegraphics[width=8cm]{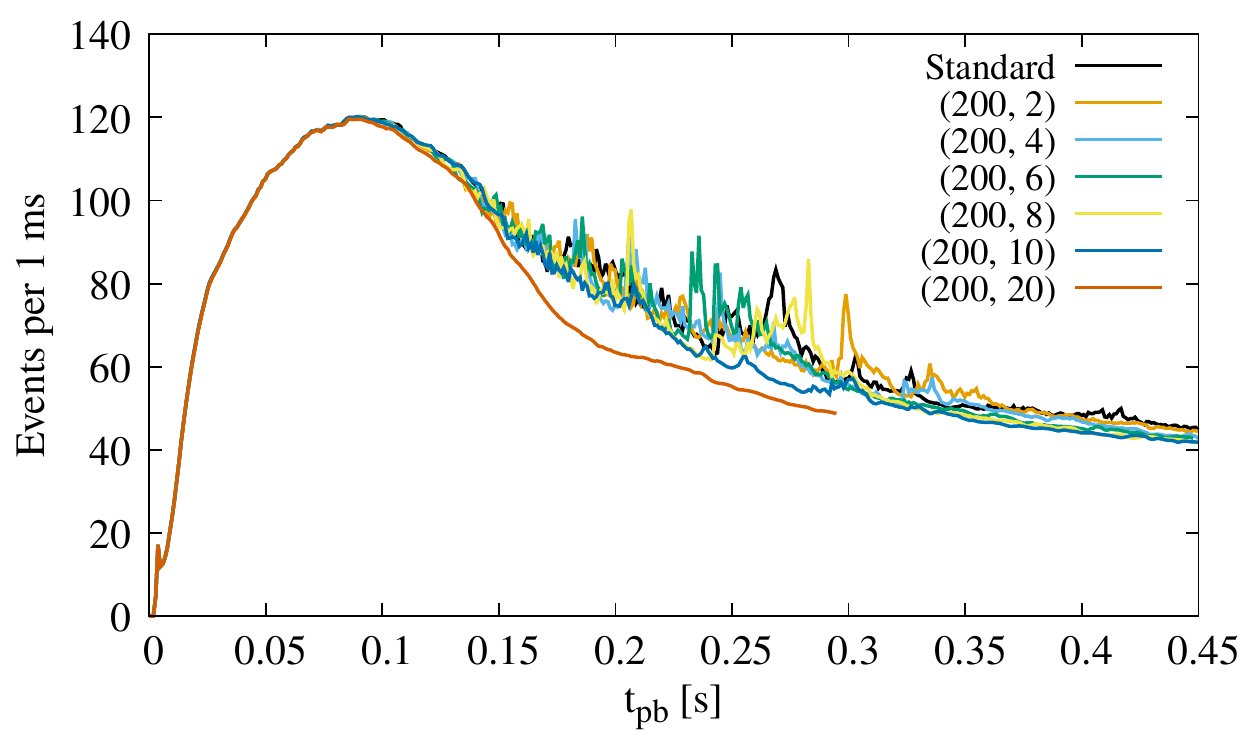}
  \caption{The same plot as Fig.~\ref{NH} but the neutrino mass hierarchy is assumed to be inverted.}
  \label{IH}
 \end{figure}

Since the stellar envelope is almost transparent to neutrinos, they can provide information on the supernova core, which is opaque to the electromagnetic waves. Figure \ref{Ln} shows the luminosity of $\nu_e$, $\bar{\nu}_e$, and $\nu_X$, where $\nu_X$ is heavy-flavor neutrinos and antineutrinos. In the case of $\nu_e$, the neutronization burst is seen soon after the core bounce when the bounce shock comes out of the neutrino sphere. Until $t_\mathrm{pb}\sim0.1$\,s, the neutrino luminosities are independent of the ALP parameters for all flavors. Except for the $(m_a/1\,\mathrm{MeV},\;g_{10})=(200,\;20)$ model, the mass accretion powers the neutrino luminosities until $t_\mathrm{pb}\sim0.2$--0.3\,s, depending on $m_a$ and $g_{10}$. After that, the shock wave is revived and the mass accretion rate drops. In this phase, neutrinos are mainly emitted from the cooling PNS. However in the $(200,\;20)$ model, the accretion stops earlier than the other models because of efficient ALP heating. As a result, the neutrino luminosities begin decreasing earlier. Because  ALP heating prevents the mass accretion, the models with larger $g_{a\gamma}$ show lower neutrino luminosities in the cooling phase.
 
Figure \ref{En} shows the mean neutrino energies. It is seen that the mean energy of heavy-flavor neutrinos, $\langle E(\nu_X)\rangle$, is larger than the mean energies of the other flavors,  because the neutrinosphere for $\nu_X$ is located at a smaller radius. Also, the mean energy of electron neutrinos, $\langle E(\nu_e)\rangle$, is smaller than the others because they react with abundant neutrons through the charged current reaction. This well-known energy hierarchy among different flavors is independent of ALPs. Also during the accretion phase, the neutrino mean energies are not affected by ALPs qualitatively. However, in the cooling phase, $\langle E(\nu_e)\rangle$ and $\langle E(\bar{\nu}_e)\rangle$ become smaller than those in the standard model because ALPs induce additional cooling of the PNS.

If a supernova event were to occur in our Galaxy, many neutrinos would be detected by terrestrial instruments. The effects of ALPs on supernova neutrinos could be imprinted in the observed signals. The number of neutrinos detected by an instruments per a unit time can be written as \cite[e.g.,][]{2016MNRAS.461.3296N,2020PhRvD.101f3027S}
 \begin{eqnarray}
     \frac{dN}{dt}=N_\mathrm{tar}\int_{E_\mathrm{th}}^\infty F(E)\sigma(E) dE,
\end{eqnarray}
where $N_\mathrm{tar}$ is the number of targets, $E_\mathrm{th}$ is the threshold energy, $F(E)$ is the number flux of neutrinos, and $\sigma(E)$ is the cross section between the targets and neutrinos. The neutrino flux is given as $F(E)=L_\mathrm{n}f(E)/4\pi D^2$, where $L_\mathrm{n}$ is the neutrino number luminosity, $f(E)$ is the neutrino distribution function, and $D$ is the distance to the supernova. The neutrino distribution can be fitted as \cite{2003ApJ...590..971K}
\begin{eqnarray}
     f(E)=\frac{(1+\alpha)^{(1+\alpha)}}{\Gamma(1+\alpha)}\frac{E^\alpha}{\langle E\rangle^{\alpha+1}}\exp\left(-(1+\alpha)\frac{E}{\langle E\rangle}\right),
\end{eqnarray}
where $\alpha=(\langle E^2\rangle-2\langle E\rangle)/(\langle E\rangle-\langle E^2\rangle)$. We adopt the cross section of the inverse $\beta$ decay ($\bar{\nu}_e+p\rightarrow e^++n$) of $\sigma(E)=9.52\times 10^{-44}(E_{e^+}p_{e^+}/1\,\mathrm{MeV}^2)$\,cm$^2$, where $E_{e^+}$ and $p_{e^+}$ are the positron energy and momentum, respectively \cite{2002RvMP...74..297B}.

When we estimate the number of neutrino events, we should consider the effects of neutrino oscillation between the source and the observer. Here, we consider the Mikheyev-Smirnov-Wolfenstein (MSW) effect \cite{1979PhRvD..20.2634W,1986NCimC...9...17M,PhysRevLett.56.1305} in the stellar envelope and vacuum oscillation, following the prescription in Ref.~\cite{2017ApJ...848...48K}. The observed flux of $\bar{\nu}_e$ is given as $F(\bar{\nu}_e)=pF^0(\bar{\nu}_e)+(1-p)F^0(\bar{\nu}_X)$, where $F^0$ is the neutrino flux emitted from the neutrinosphere. The survival probability $p$ is given by $p=\cos^2\theta_{12}\cos^2\theta_{13}\approx0.676$ for the normal mass hierarchy and $p=\sin^2\theta_{13}\approx0.0234$ for the inverted mass hierarchy. The Earth effect is not taken into account.
 
In the case of Hyper-Kamiokande (HK), the number of target protons can be estimated as
$N_\mathrm{tar}=N_\mathrm{A}(2M_\mathrm{H}/M_\mathrm{H_2O})\rho_\mathrm{H_2O}V$,
where $N_\mathrm{A}$ is the Avogadro constant, $2M_\mathrm{H}/M_\mathrm{H_2O}=2/18$ is the mass fraction of protons in a water molecule, $\rho_\mathrm{H_2O}$ is the water density, and $V=220$\,kton is the detector volume \cite{2017ApJ...848...48K,2018arXiv180504163H}. The threshold energy is set to $E_\mathrm{th}=8.3$\,MeV \cite{2017ApJ...848...48K} and we assume that the detection efficiency is 100\% in $E>E_\mathrm{th}$. Figures \ref{NH} and \ref{IH} show the number of neutrino events expected at HK, assuming a supernova event at the Galactic center (i.e., $D=8.5$\,kpc). Figure \ref{NH} assumes the normal mass hierarchy, while Fig.~\ref{IH} assumes the inverted mass hierarchy. Regardless of the ALP parameters, HK would detect $\sim120$ events per 1\,ms at the peak. It is notable that the rise time is shorter in the case of the inverted mass hierarchy. This implies that the early phase of the $\bar{\nu}_e$ signals is useful to determine the neutrino mass hierarchy \cite{2012PhRvD..85h5031S}. During the accretion phase, the signal would be stochastic and dependence on the ALP parameters is not clear. In the PNS cooling phase, the signal becomes more smooth and the event number tends to be smaller with larger $g_{a\gamma}$, although dependence on $g_{a\gamma}$ is not monotonic at this stage. Dependence on $g_{a\gamma}$ is more significant in the case of the normal hierarchy than the case of the inverted hierarchy. This is because the $\bar{\nu}_X$ luminosity and mean energy are not very sensitive to ALPs, and the $\bar{\nu}_e$ flux on Earth is mainly determined by the $\bar{\nu}_X$ flux in the case of the inverted hierarchy because of the MSW effect.

\subsection{Gravitational Waves}
    \begin{figure}
  \centering
  \includegraphics[width=8cm]{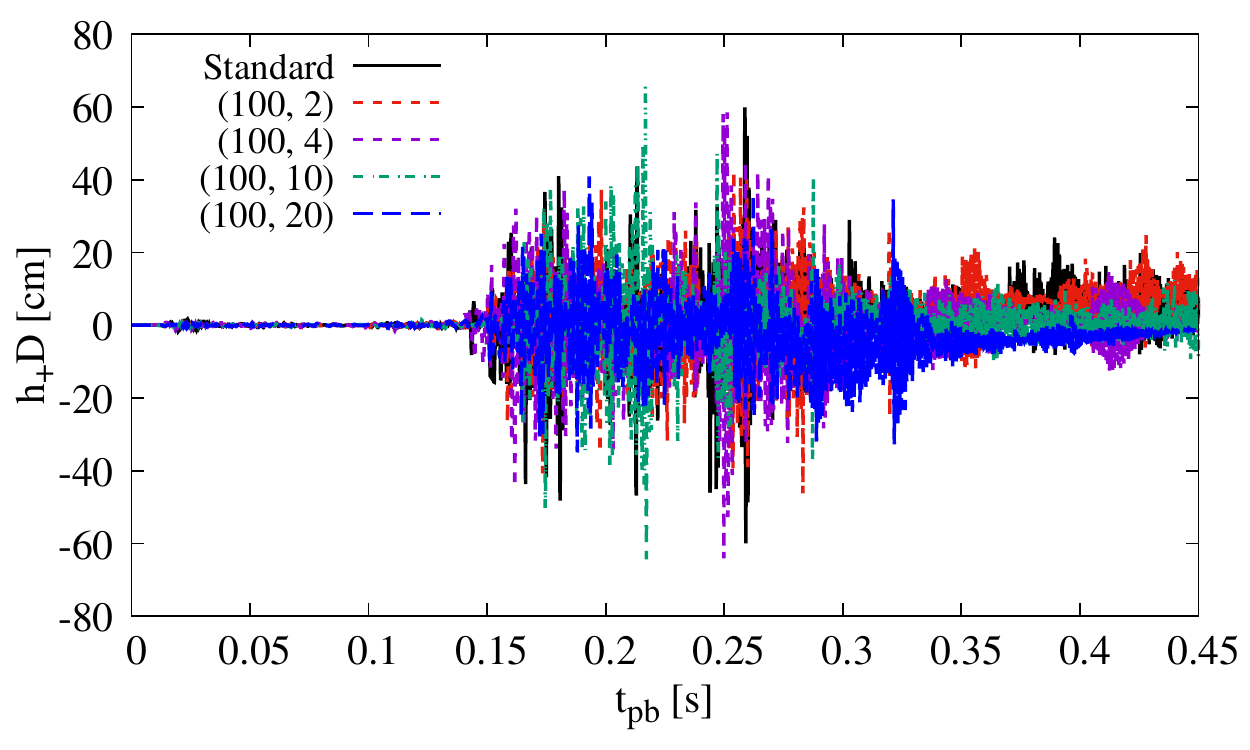}
   \includegraphics[width=8cm]{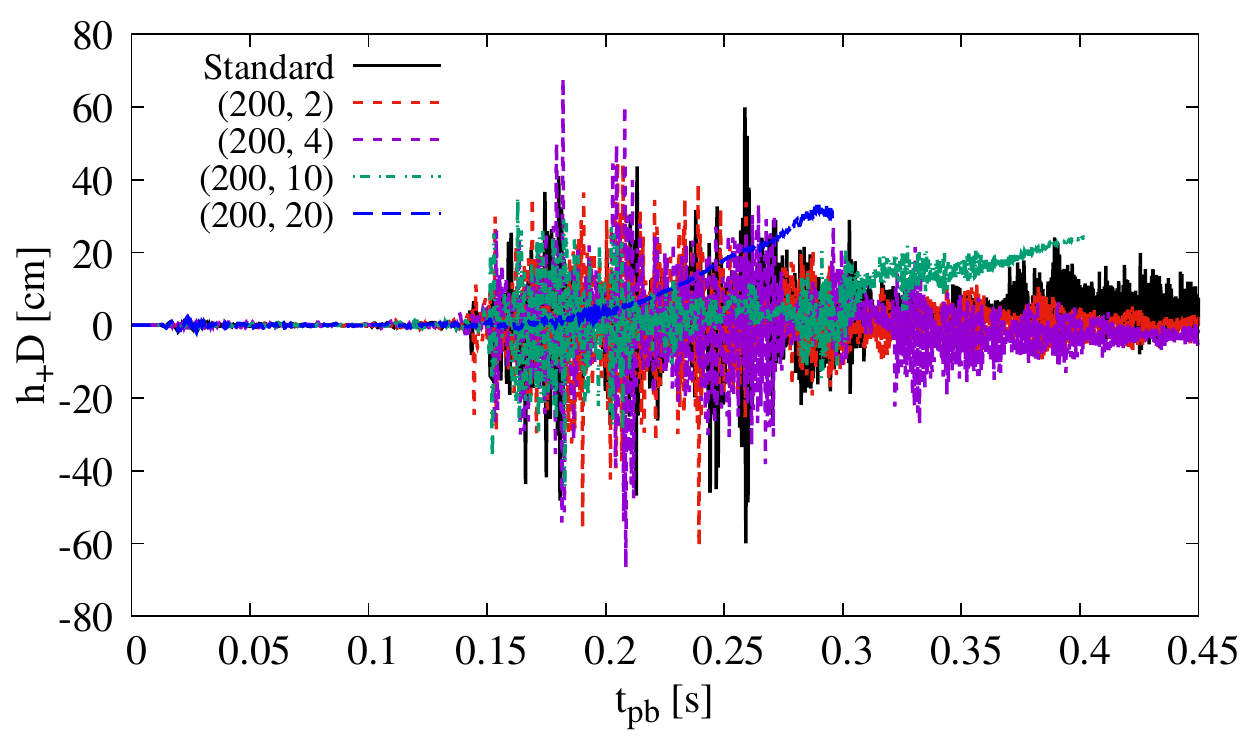}
  \caption{The GW strain $h_+$ times the distance $D$ to the event in the selected models as a function of time $t_\mathrm{pb}$ after the core bounce. The upper panel shows the results for the models with $m_a=100$\,MeV and the lower panel is for the models with $m_a=200$\,MeV. The solid line corresponds to the model without ALPs, and the other lines correspond to the models with ALPs.}
  \label{GW}
 \end{figure}

     \begin{figure}
  \centering
  \includegraphics[width=8cm]{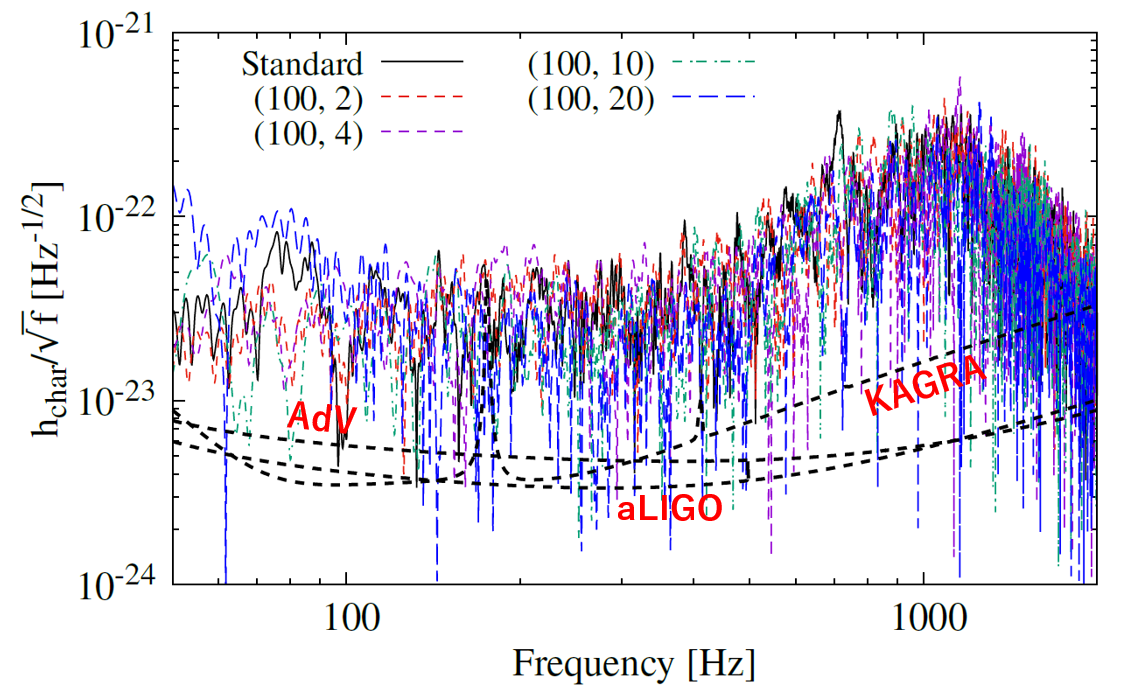}
   \includegraphics[width=8cm]{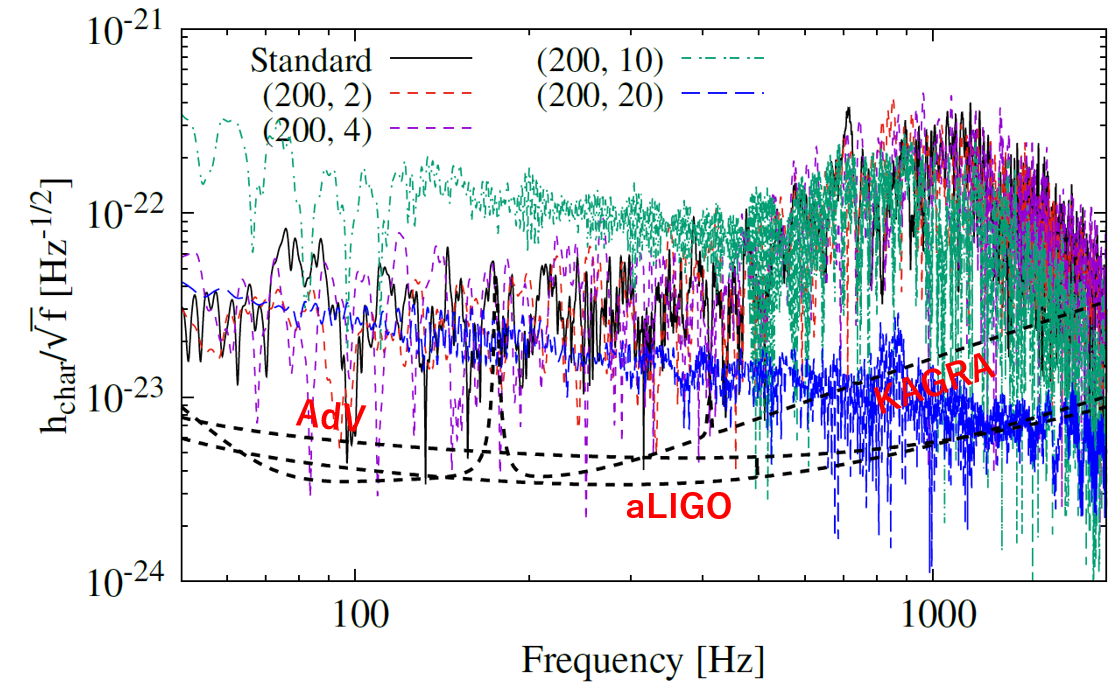}
  \caption{The GW characteristic strain $h_\mathrm{char}$ divided by $\sqrt{f}$ in the selected models, where $f$ is the GW frequency. The sensitivity of Advanced LIGO \cite[aLIGO;][]{aLIGO}, Advanced VIRGO \cite[AdV;][]{AdV}, and KAGRA \cite{KAGRA} is also shown. The upper panel shows the results for the models with $m_a=100$\,MeV and the lower panel is for the models with $m_a=200$\,MeV.}
  \label{GW_spec}
 \end{figure}
If the explosion deviates from spherical symmetry, the system would produce GWs. Because our models are axisymmetric, only the plus mode of GWs is tracked. The GW strain $h_+$ is given as
\begin{eqnarray}
    h_+=\frac{3}{2}\frac{G}{Dc^4}\sin^4\alpha\frac{d^2}{dt^2}\ibar_{zz},
\end{eqnarray}
where $\alpha$ is the angle between the line of sight and the symmetry axis, and $\ibar_{zz}$ is the only independent component of the reduced quadrupole moment. The first time derivative of $\ibar_{zz}$ can be evaluated as \cite{2009ApJ...707.1173M,2016MNRAS.461.3296N}
\begin{equation}
\begin{split}
    \frac{d}{dt}\ibar_{zz}&=\frac{8\pi}{3}\int d\cos\theta\int dr r^3\rho\times \\
    &\left(P_2(\cos\theta)v_r+\frac{1}{2}\frac{\partial}{\partial\theta}P_2(\cos\theta)v_\theta\right),
\end{split}
\end{equation}
where $v_r$ and $v_\theta$ are the radial and lateral velocities, $\rho$ is the density, and $P_2(x)$ is the second Legendre polynomial. In this section, we calculate GW signals observed on Earth assuming $\sin\alpha=1$.

Figure \ref{GW} shows the GW strain as a function of post-bounce time $t_\mathrm{pb}$. In all of the models, the GW signals are quiet before $t_\mathrm{pb}\approx0.15$\,s because the core is approximately spherically symmetric at this stage. After this time, a SASI-like instability starts and GWs are produced. The GW waveform in the standard model and the models with small $g_{a\gamma}$ is similar to those in non-rotational two-dimensional models without magnetic fields reported in previous works \cite[e.g.,][]{2007ApJ...655..406K,2009ApJ...707.1173M}.

In Fig.~\ref{GW}, we can see that $h_+$ becomes smaller if ALPs are considered. This is because the ALP heating suppresses the mass accretion on the PNS. In the models with $(m_a/1\,\mathrm{MeV},\;g_{10})=(200,\;10)$ and $(200,\;20)$, $h_+$ becomes positive with a low frequency. In these models, the strong ALP heating induces prolate explosion in the early stages. If we continued the simulations out to longer times and the morphology of explosion deviates from spherical symmetry, the other models could show similar trends too \cite{2009ApJ...707.1173M}. However, we note that most of the 2D models 
 lead to the prolate explosion toward the 2D coordinate symmetry axis \cite[e.g.][]{2013CRPhy..14..318K}. The explosion morphology can be more anisotropic in three-dimensional models \cite{2019ApJ...876L...9R,2022MNRAS.514.3941N}, which are beyond the scope of this work.

One can Fourier-transform the time series of $h_+$ to obtain the GW spectral energy distribution. It is common to define the characteristic strain \cite{PhysRevD.57.4535}
\begin{eqnarray}
    h_\mathrm{char}=\sqrt{\frac{2G}{\pi c^3D^2}\frac{dE_\mathrm{GW}}{df}},
\end{eqnarray}
where $dE_\mathrm{GW}/df$ is the GW spectral energy density. Figure \ref{GW_spec} shows $h_\mathrm{char}$ in our models with $D=8.5$\;kpc. In all of the models except for $(m_a/1\,\mathrm{MeV},\;g_{10})=(200,\;20)$, we can find a broad peak at $f\sim1$\,kHz. This is attributed to the SASI-like motion that operates at $t\gtrsim0.15$\,s \cite[e.g.,][]{2009ApJ...707.1173M}. In the models with (200,\;10) and (200,\;20), $h_\mathrm{char}/\sqrt{f}$ increases toward a low frequency. This low-frequency feature is also seen in Fig.~\ref{GW} and comes from the prolate explosion. The figure also indicates the sensitivity of Advanced LIGO \cite{aLIGO}, Advanced VIRGO \cite{AdV}, and KAGRA \cite{KAGRA}. The comparison between the model prediction and the instrumental sensitivity implies that the GW detectors would detect the GW signals from a supernova event at the Galactic center. It is seen that the strain in the model with $(200,\;20)$, in which the ALP heating is most efficient, is lower than the others by $\sim10$ times. It is hence more difficult to detect GWs if the ALP heating is too effective, although even this model predicts the high GW amplitude enough to be detected.


\section{Discussion and Conclusions}
In this study, we performed two-dimensional axisymmetric supernova simulations which consider the effects of the ALP-photon interaction. It was found that heavy ALPs with $m_a=100$--200\,MeV can increase the diagnostic energy of explosion. In particular, our model with $m_a=200$\,MeV and $g_{10}=6$ showed $E_\mathrm{diag}\approx0.6\times10^{51}$\,erg at the end of the simulation, which is close to the observational values, while the reference model without ALPs resulted in only $E_\mathrm{diag}\approx0.4\times10^{51}$\,erg.
 
In this context, most of the recent multi-dimensional supernova simulations exhibit $E_\mathrm{diag}<0.6\times10^{51}$\,erg \cite{2015ApJ...807L..31L,2019MNRAS.489..641M,2020MNRAS.491.2715B,2022MNRAS.514.3941N,2022MNRAS.516.1752M}, which are smaller than the typical values for observed type II supernovae \cite{2022A&A...660A..41M}. The lower predicted energetics indicates that current supernova models could be lacking some physical processes that should be considered. Recently, Ref.~\cite{2021ApJ...915...28B} performed long-term three-dimensional simulations of the core-collapse of a $18.88M_\odot$ star until $t_\mathrm{pb}=7$\,s. In their models, $E_\mathrm{diag}$ continued to increase even in $t_\mathrm{pb}>1$\,s and it finally reached $0.9779\times10^{51}$\,erg at the end of the simulation. This result shows that performing long-term simulation is important to estimate the final explosion energy. However, there are still uncertainties in physical input such as the equation of state and perturbations in progenitors, and it is unclear if $10^{51}$\,erg explosion can be achieved with different inputs. Our results suggest that heavy ALPs could help supernova models reproduce $0.6\times10^{51}$\,erg explosion if $m_a\approx200$\,MeV and $g_{10}\approx6$, although the parameter would be excluded by the comparison with low-energy supernovae \cite{2022PhRvL.128v1103C}. In addition, heavy ALPs which we are focusing on can cause the cosmological inflation if they exist \cite{2019JHEP...07..095T}. It is hence important to pursue heavy ALPs in both of the astrophysical and cosmological contexts.

If ALPs are produced in a supernova core, they can affect the neutrino and GW signals from nearby events. We found that the multi-messenger signals become weaker if $g_{a\gamma}$ is high enough. In the parameter region we explored in this study, we would detect neutrinos and GWs from a supernova event at the Galactic center. We note that the systematic behavior of explosion dynamics in terms of the ALP-photon coupling constant is not monotonic when $g_{10}\lesssim10$ because of stochasticity.

We found that the neutrino signals from the supernova model with $(m_a/1\,\mathrm{MeV},\;g_{10})=(200,\;20)$ attenuate much faster than the other models because of the higher cooling rate. Also, the GW amplitude calculated in the models with $(200,\;10)$ and $(200,\;20)$ is significantly suppressed compared with the other models with weaker ALP heating. Although these differences would lead to detectable signatures in observed data, these ALP parameters result in the diagnostic explosion energy higher than $2\times10^{51}$\,erg, which is more energetic than typical type II supernovae. This implies that, whereas the explosion energy is a useful observable to study supernova ALPs, ALPs with parameters allowed by the explosion energy argument \cite{2022PhRvL.128v1103C} are not likely to impact GW and neutrino observations of a nearby supernova event.

In our simulations, the calculation is stopped at $t_\mathrm{pb}\sim0.5$~s. This is long enough to study the core bounce and the mass accretion phase in detail, but it is desirable to perform long-term simulations \cite{2016PhRvD..94h5012F,2021PhRvD.104j3012F,2022arXiv220914318F} to predict neutrino counts in detectors from nearby supernovae. The signals from SN 1987A lasted for $\sim10$\,s \cite{1987PhRvL..58.1494B,1987PhRvL..58.1490H,1987JETPL..45..589A} and currently operating detectors will follow the neutrino emission for 10--100 seconds. Long-term simulations are important also because the ALP energy deposition $E_a$ in the outer layers would significantly enhance the asymptotic value of the explosion energy, as mentioned in Section III.A. Additionally, two-dimensional models can predict only the plus mode of GWs because they assume axisymmetry. It is necessary to develop three-dimensional models to predict every mode of GWs observed from arbitrary direction. 

Although the ALP parameter region explored in this study is marginally excluded by comparison with low-energy supernovae \cite{2022PhRvL.128v1103C}, we can use the methodology developed here to investigate any exotic feebly-interacting particles such as sterile neutrinos \cite{2014PhRvD..90j3007W,PhysRevD.98.103010}. It is desirable to perform core-collapse simulations with the transport of various exotic particles and predict multi-messenger signals before the next nearby supernova event appears in order to fully use the event as a laboratory of new physics.

\begin{acknowledgments}
K.M. is grateful to Ko Nakamura for stimulating discussions. Numerical computations were  carried out on Cray XC50 at Center for Computational Astrophysics, National Astronomical Observatory of Japan. This work is supported by Research Institute of Stellar Explosive Phenomena at Fukuoka University and the University Project No. GR2302, and JSPS KAKENHI Grant Numbers JP21K20369, JP17H06364, JP18H01212, JP21H01088, JP22H01223, JP23KJ2147, JP23K03400, 23H01199 and JP23K13107. 
The work of SH is supported by the U.S.~Department of Energy Office of Science under award number DE-SC0020262, NSF Grant No.~AST1908960 and No.~PHY-1914409 and No.~PHY-2209420, and JSPS KAKENHI Grant Number JP22K03630 and JP23H04899. This work was supported by World Premier International Research Center Initiative (WPI Initiative), MEXT, Japan.
\end{acknowledgments}

\bibliography{ref.bib}
\end{document}